# Pressure-Driven Magneto-Topological Phase Transition in a magnetic Weyl semimetal

Qingqi Zeng[1†], Hongyi Sun[2†], Jianlei Shen[1], Qiushi Yao[3], Qian Zhang[4], Nana Li[4], Lin Jiao[5], Hongxiang Wei[1], Claudia Felser[6], Yonggang Wang[4*], Qihang Liu[3,7*], Enke Liu[1,8*]

1. State Key Laboratory for Magnetism, Beijing National Laboratory for Condensed Matter Physics, Institute of Physics, Chinese Academy of Sciences, Beijing 100190, China

2. Shenzhen Institute for Quantum Science and Engineering, Southern University of Science and Technology, Shenzhen 518055, China

3. Shenzhen Institute for Quantum Science and Technology and Department of Physics, Southern University of Science and Technology (SUSTech), Shenzhen 518055, China

4. Center for High Pressure Science and Technology Advanced Research (HPSTAR), Beijing 100094, China

5. Department of Physics and Frederick Seitz Materials Research Laboratory, University of Illinois Urbana-Champaign, Urbana, Illinois 61801, USA

6. Max Planck Institute for Chemical Physics of Solids, Dresden 01069, Germany

7. Guangdong Provincial Key Laboratory for Computational Science and Material Design, Southern University of Science and Technology, Shenzhen 518055, China

8. Songshan Lake Materials Laboratory, Dongguan, Guangdong 523808, China

---

[†] These authors contributed equally to this work.
[*] ekliu@iphy.ac.cn (E. L.)
[*] liuqh@sustech.edu.cn (Q. L.)
[*] yonggang.wang@hpstar.ac.cn (Y. W.)

# Abstract

The co-occurrence of phase transitions with local and global order parameters, such as the entangled magnetization and topological invariant, is attractive but has been seldom realized experimentally. Here, by using high-pressure *in-situ* X-ray diffraction, high-pressure electric transport measurements and high-pressure first-principles calculations, we report a magneto-topological phase transition, i.e., the phenomenon of magnetic materials undergoing different magnetic and topological phases during the process of pressure loading, in a recently discovered magnetic Weyl semimetal $Co_3Sn_2S_2$. By considering both out-of-plane ferromagnetic and in-plane anti-ferromagnetic components, the calculated results can well fit the experimental data. The calculation results furtherly reveal a pristine Weyl phase with four more pairs of Weyl nodes under low pressures, and a generally-defined $Z_2$ topological insulator phase after the restoration of time-reversal symmetry. Remarkably, the present magneto-topological phase transition involves a pair of crossing bands of two spin channels becoming degenerate. Thus, all the chiral Weyl nodes annihilate with their counterparts from another spin channel, in contrast to the typical annihilation of Weyl pairs from the same bands in inversion-asymmetric systems. Our experiments and theoretical calculations uncover a manner to modulate the diverse topological states by controlling the internal exchange splitting via external physical knobs in topological magnets.

**Introduction**

The topological phase transition (TPT) is a class of phase transition beyond the Landau paradigm, which cannot be described by spontaneous breaking of a local symmetry[1-4]. As a classical case, the phase transition from a normal insulator to a topological insulator (TI) connects radically different topological electronic characteristics of two insulating phases[5]. In contrast, the evolution between Dirac fermions with four-fold band degeneracy and chiral Weyl fermions with two-fold degeneracy can be realized between two metallic phases by breaking either the inversion symmetry or the time-reversal symmetry (TRS)[5-6]. In particular, the TPT by changing TRS triggers the interplay between topology and magnetism[7-8]. When a TPT is accompanied or even induced by a magnetic phase transition, the entangled degrees of freedom can be tuned cooperatively by multiple conditions (such as temperature and pressure), which is attractive to the realization of various topological quasiparticles[9-10] and to the promising applications of novel topological spintronics[11-14].

We here consider the emerging Weyl-TI phase transition, which contains two distinct topological nontrivial phases as well as metal-insulator transition. If TRS is preserved, the chiral Weyl nodes annihilate in pairs with each other upon phase transition (see Figure 1(a)). However, it is difficult to find a generic tuning parameter to create the non-diagonal term of the Weyl Hamiltonian and thus trigger the TI phase. In contrast, when magnetism is involved, the magnetic order naturally performs as an order parameter driving the material to different topological phases during the spontaneous magnetization. It is desired that the Weyl system could become a $Z_2$ topological insulator if the band inversion persists when the magnetism is totally suppressed by pressure or temperature. In addition, such magneto-TPT will reveal a new type of Weyl-nodes annihilation. As shown in Figure 1(b), the Weyl nodes from one spin channel could meet and annihilate with the corresponding band crossings of the other spin channel. However, realizing such magnetism-involved TPT is technically challenging and yet to be validated experimentally, because it highlights the essence of intrinsic topological magnets that are rarely explored. Recently, TPT around the Curie temperature of $Co_3Sn_2S_2$ is revealed by temperature-dependent Angle resolved photoemission spectroscopy (ARPES) studies [15-17], which indicates that the transition from Weyl semimetal to TI in $Co_3Sn_2S_2$ can be driven by temperature. Compared with the thermal effect that destroys magnetic order at relative high temperatures, pressure can suppress the magnetism at low temperatures and thus should observe the intrinsic physical behaviors more purely.

In this work, we focus on the TPT and apply high pressure to study the evolution of the topological states in the Weyl magnet $Co_3Sn_2S_2$. Since its discovery [18-21], remarkable topological signatures, including giant anomalous Hall (AHE) [18-19], anomalous Nernst effects [22-23], surface Fermi arcs [20-21], chiral edge states [24-25], giant magneto-optical Kerr effect[26] and spin-orbit polaron[27], have been observed in $Co_3Sn_2S_2$, rendering it an ideal platform for TPT with TRS breaking and restoring. Recently, pressure studies have been performed to tune the giant AHE of $Co_3Sn_2S_2$, indicating the intrinsic-mechanism-dominated AHE decreases during the loading pressure[28-29]. The pressure can suppress the ferromagnetism and reduce the distance between the Weyl nodes with opposite chirality. However, the potential TPT as well as the evolution of topological state upon the phase transition remains unaddressed.

Combining transport measurements and first-principles calculations, we find that with increasing pressure, the out-of-plane ferromagnetic (FM) ordering is suppressed while in-plane anti-ferromagnetic (AFM) components arises in $Co_3Sn_2S_2$, with a concurrent decreased intrinsic anomalous Hall conductivity (AHC). From ambient pressure up to ~40 GPa, the system evolves from a low-pressure Weyl phase to a high-pressure Weyl phase, and finally to a general TI phase with Pauli paramagnetism (PM), accompanied with vanishing spin polarization and AHC. Unlike the pressure-induced TPT in nonmagnetic materials where the pressure triggers band inversion [30-31], the spontaneous magnetism plays a role as a pivot that drives the TPT in response to pressure. Furthermore, since our magneto-TPT involves the bands of two spin channels becoming degenerate by restoring the TRS, all of the Weyl nodes annihilate with their counterparts from another spin channel in higher conduction bands. Hence, a generic magneto-TPT with new mechanism of Weyl-nodes annihilation is found in this topological magnet by tuning well-controlled parameters such as pressure and temperature (see Figure 1(c)).

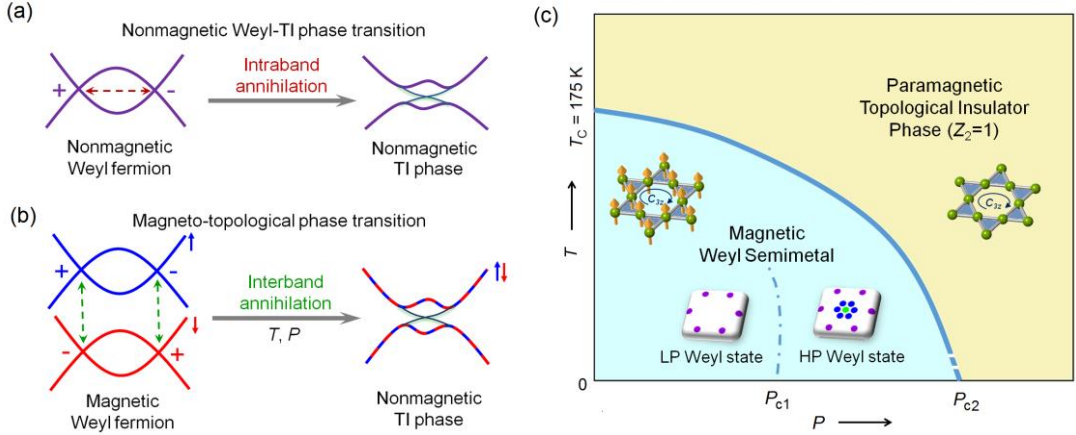

**Figure 1.** Topological phase transition (TPT). (a) Weyl-TI TPT with the intraband annihilation of chiral Weyl nodes in a nonmagnetic system (in this case the inversion symmetry should be broken, e.g., TaAs). (b) Weyl-TI TPT driven by temperature or pressure with the interband annihilation of chiral Weyl nodes in a magnetic system. The dashed arrows denote the trajectory of the Weyl-node annihilation (in this case the inversion symmetry is preserved, e.g., $Co_3Sn_2S_2$). For generality, we assume that the phase transition preserves the crystal symmetry. (c). Magneto-topological phase diagram of $Co_3Sn_2S_2$ in the space of pressure ($P$) and temperature ($T$). The Kagome lattices with and without ferromagnetic order correspond to the magnetic Weyl semimetal and topological insulator phase, respectively. $T_C$ is the Curie temperature. $P_{c1}$ and $P_{c2}$ stand for the critical pressures of the first and second TPTs, respectively. Low-pressure (LP) Weyl phase contains 6 Weyl nodes (purple spots), while high-pressure (HP) Weyl phase contains 8 more, including 6 connected by inversion and $C_{3z}$-rotation symmetries (blue spots), and 2 located at the $k_z$ axis of the Brillouin zone ($k_x = k_y = 0$, green spots).

**Result and discussion**

The Shandite $Co_3Sn_2S_2$ adopts a rhombohedral structure with the *R-3m* space group. The unit cell in a hexagonal representation is shown in Figure 2(a), in which one Sn atom is in the center of the Kagome lattice formed by Co atoms, and S and the remaining Sn atoms are located between the $Co_3Sn$ Kagome layers. $Co_3Sn_2S_2$ is known to be an itinerant ferromagnet with magnetic moment 0.29 $\mu_B$ per Co atom [18, 32], with the measured easy axis along the *c* axis [33], as indicated by the red arrows in Figure 2(b). A schematic sketch of the band inversion of Co orbitals on Kagome lattice is also shown in Figure 2(b). The red and blue spots show the Weyl nodes with opposite chirality, which act as monopole sink and source of Berry curvature in momentum space [18]. The surface

Fermi arc connects the projection of two Weyl nodes [18, 20-21]. All these topological characters exist on the Co-atom Kagome lattice in $Co_3Sn_2S_2$. Pressure-dependent studies of $Co_3Sn_2S_2$ were conducted using a Be/Cu diamond anvil cell (DAC) with X-ray and magnetic field passing through the diamond windows (as shown in Figure 2(c)). The $Co_3Sn_2S_2$ samples used in this work were single-crystalline microflakes (see details in *Methods* and Supplemental Materials Note 1). Figure 2(d) shows the XRD patterns of $Co_3Sn_2S_2$ under pressures, from which no obvious structural phase transition was observed up to 40 GPa. All the XRD data were thus refined with space group *R*-3*m* (as shown in Figure 2(e) and Supplemental Materials Note 2). The lattice parameters *a* and *c* along with the cell volume *V* decrease with increasing pressure, which is also observed in previous report[28]. Both the low-pressure region (0 - 10 GPa) and high-pressure region (15 - 40 GPa) can be well fitted using the third-order Birch-Murnaghan equation of state, respectively [34]. Between 10 and 15 GPa, the compression of the unit cell seems abating, which may be due to a subtle electronic transition.

We measured the temperature dependence of longitudinal resistance (*R-T*) of $Co_3Sn_2S_2$ under high pressures, which are shown in Figure 2(f). All first-order differential of *R-T* curves were further recorded and the abrupt changes (kink point on *R-T* curves) of which were determined to be Curie temperature $T_C$ (see Supplemental Materials Note 3 for details). The $T_C$ decreases with increasing pressure (the data from 0.5 to 17 GPa are shown as solid symbols in Figure 2(g)), similar to the observations at previous studies[28-29, 35]. There are kinks at about 45 K, which remain almost unchanged above 17 GPa (shown as open symbols in Figure 2(g)). According to the analysis about the coercive field (illustrated below), the magnetism in this system is indeed suppressed by high pressures. Hence, the unchanged kink is not attributed to $T_C$ above 17 GPa in this text. There may be other transition in this system, which results in the unchanged kink temperature ($T_k$) above 17 GPa.

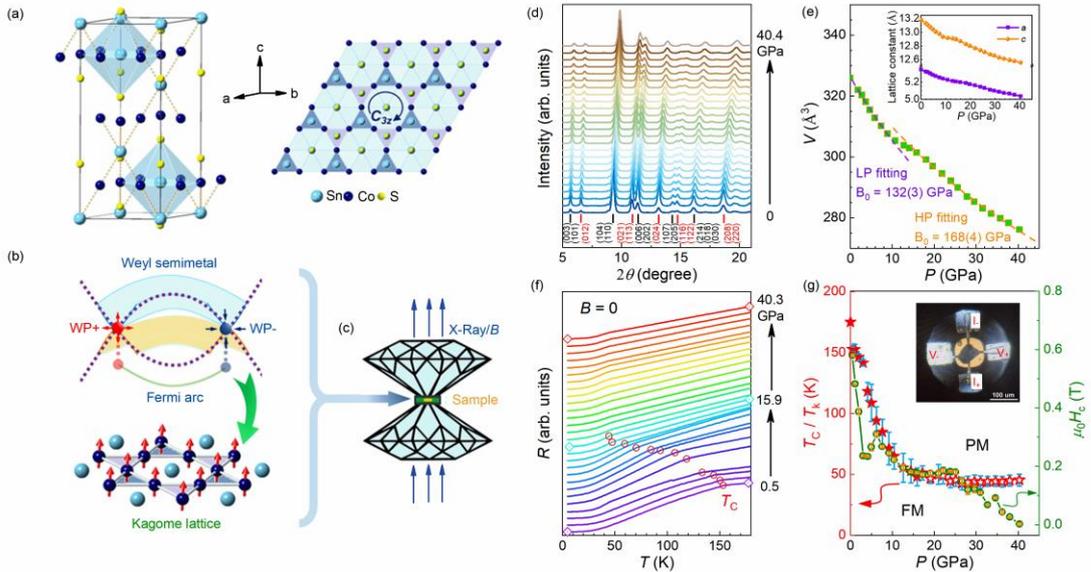

**Figure 2.** Crystal structure, topological state, pressure-dependent X-ray diffraction and electric resistivity. (a) Hexagonal unit cell of $Co_3Sn_2S_2$. Kagome lattice formed by Co atoms are depicted in the right panel of (a) and lower panel of (b) in different perspectives. The magnetic moments in Co sites which are along *c* axis direction are indicated by red arrows. A sketch of band inversion, Fermi arc and Weyl nodes on Co-atom Kagome lattice is shown in the upper panel of (b). (c) Sketch of high-pressure experimental configuration. (d) X-ray diffraction patterns of $Co_3Sn_2S_2$ under different

pressures. (e) Pressure dependence of unit cell volume in a rhombohedral space group *R*-3*m*. Purple and orange dash lines are fitting results based on the equation of state. Inset shows the pressure dependence of lattice parameters. (f) Temperature dependence of the resistance in arbitrary unit under different pressures. The diamond symbols are the guides of the eyes, through which one can easily find the curves under 0.5, 15.9 and 40.3 GPa. (g) Pressure dependences of Curie temperature (solid star), kink temperature (open star) and coercive field. "PM" and "FM" indicate the paramagnetic and ferromagnetic states, respectively. Inset shows the sample and configuration of Hall measurements.

As the coercive field in magnetization loop can also be an indicator of the FM order, we further made analysis of the coercive field ($H_c$) extracted from Hall data measured at 5 K (see Figure 2(g) and Supplemental Materials Note 3.3 for details). It reveals that the FM order changes little between 17 to 25 GPa, but decreases rapidly above 26 GPa and nearly vanishes around 40 GPa. It can thus be concluded that the itinerant FM order approaches near non-magnetic state, or in other words, Pauli PM[36], above the critical pressure of 40 GPa.

Figure 3(a–c) show the Hall data under different pressures. At a glance of Figure 3(a), the data show a linear, positive-slope behavior for high-field normal Hall part. It shows the dominance of hole carriers in this system. One can see that the transverse resistivity at zero-field $\rho_{yx}(0)$, i. e., the anomalous Hall resistivity rapidly decreases with increasing pressure, which is also shown in the inset of Figure 3(d). With an amplified scale (Figure 3(b)), the Hall curves show positive slopes below 26.6 GPa, above which the slope changes to negative, indicating a switch of the dominant carrier type. In the two previous reports, the switch of carrier type are observed at lower pressure[28] and not observed[29], respectively. Above-mentioned differences may be due to the variation of samples. Meanwhile, the loop-hysteresis character gradually becomes faint (Figure 3(c)), leaving a nearly linear response of normal Hall effect at high pressures. At 40 GPa, both the AHE and $H_c$ finally approach zero (also see Figure 3(d) and 2(g)). Correspondingly, the AHC decreases remarkably from 1550 $\Omega^{-1}$ cm$^{-1}$ to nearly zero with increasing pressure up to 40 GPa (Figure 3(d)), which uncovers the near vanishment of FM order under high pressures.

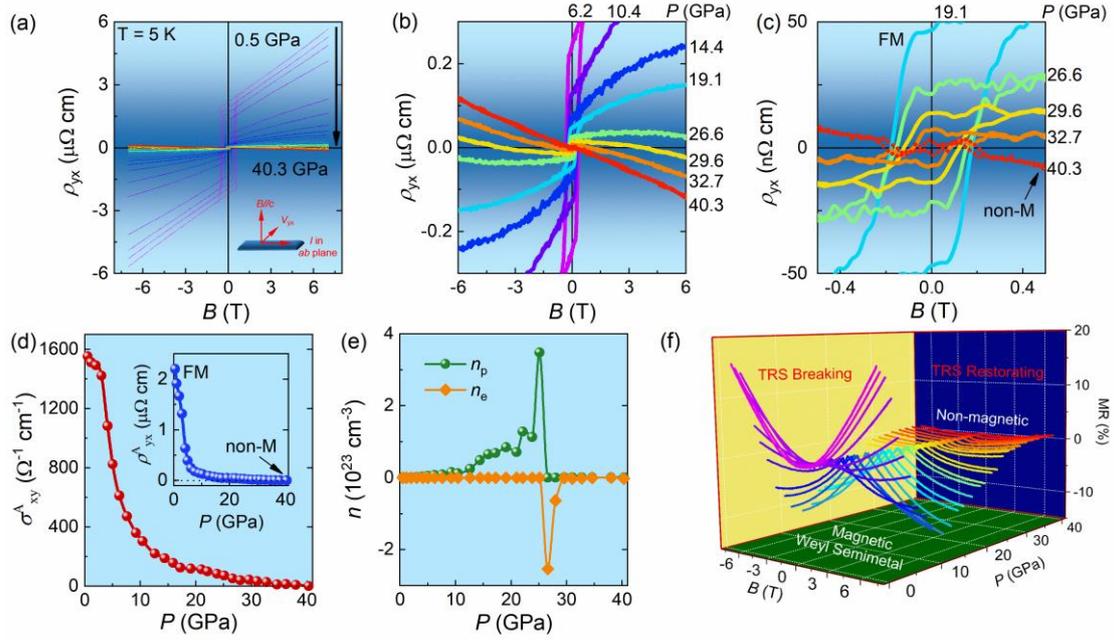

**Figure 3.** Hall effect, carrier, and MR under pressure. (a), (b) and (c) Magnetic field dependence of Hall resistivity under different pressures. (b, c) are enlarged scales of (a). (d) Pressure dependence of the anomalous Hall conductivity. Inset shows the pressure dependence of anomalous Hall resistivity. (e) Pressure dependences of carrier concentrations. The subscript p and e denote holes and electrons respectively in two-band analysis. (f) Pressure dependence of MR at 5 K. At low pressures, the system shows magnetic Weyl state with large MR; while at high pressures the system becomes non-magnetic with small MRs.

As the Hall curves show an observable nonlinear behavior in high-field region between the pressures of 10 and 30 GPa (Figure 3(b)), we employed the two-band model [37-38] on all the Hall data measured at 5 K (see *Methods* and Supplemental Materials Note 3.4 for details) to get the carrier concentrations (shown in Figure 3(e)). During the pressure loading, one can see a notable switch of the carrier types around 25 ~ 27 GPa. Below 25.1 GPa the transport behavior is dominated by holes, with the hole concentration increasing from $8.5 \times 10^{20}$ cm$^{-3}$ ($P = 0.5$ GPa) to $3.5 \times 10^{23}$ cm$^{-3}$ ($P = 25.1$ GPa), by three orders in magnitude. Above 26.6 GPa, the dominant carrier becomes electrons rapidly. The electron concentration also shows a maximum value of $2.5 \times 10^{23}$ cm$^{-3}$ ($P = 26.6$ GPa) and decreases to $2.9 \times 10^{21}$ cm$^{-3}$ ($P = 40.3$ GPa), by two orders in magnitude, leaving a relatively high concentration of electrons. During this process it can be known that the Fermi surfaces of electrons and holes increase simultaneously and a compensation between two carriers with high concentrations occurs at a pressure close to 26 GPa. Recall that a synchronous sudden decrease of $H_c$ occurs around 26 GPa (Figure 2(g)), there may exist a strong correlation between the evolutions of electronic structure and spin splitting under pressure.

Figure 3(f) exhibits the pressure dependence of the magnetoresistance (MR). Below 3.0 GPa, a positive MR is observed, which is consistent with that of bulk samples under zero pressure [18]. With increasing pressure, the MR decreases rapidly, finally to negative values above 6.2 GPa, showing a maximum around 12.6 GPa. Negative MR appears usually in magnetic materials due to the spin-related scattering[39]. Thus, during the loading a competition exists between the positive MR from ordinary MR and the negative one from spin-related scattering. Above 26 GPa the negative

MR decays as the ferromagnetism weakens above this pressure (also see Figure 2(g)).

We further notice that there are pronounced kinks around 3 GPa in both pressure dependence of the $T_C$, $H_c$ and AHC curves, after which the values of all above-mentioned magnetic related parameters decrease fast. According to the magnetic measurements (see Supplemental Materials Note 1) about $Co_3Sn_2S_2$ at ambient pressure, there is a magnetic transition below Curie temperature, which is also reported in previous researches[18, 35, 40-41]. Pronounced difference between the zero-field-cooling and field-cooling curves also indicate other magnetic interactions besides the out-of-plane FM one[18, 40-41] may exists. An earlier report shows there are in-plane AFM components in $Co_3Sn_2S_2$ above 90 K at ambient pressure. Furthermore, high pressure will support this magnetic phase to lower temperatures[35]. Hence, one possible reason that is responsible for the fast decreases at 3 GPa is the in-plane AFM interaction. There may be in-plane AFM components that will result in much lower AHC as the out-of-plane ferromagnetic order is crucial to the topological transport properties in this system[18-19, 35].

Nevertheless, we here prefer to stress that no matter how the magnetic order varies during the pressure loading, the studied sample approaches near non-magnetic state above 40 GPa, which can be recognized from the nearly vanished $H_c$ and AHC. Hence, one can expect a magnetic phase transition from the FM state to PM state above 40 GPa at low temperatures in $Co_3Sn_2S_2$. This means that the TRS-broken Weyl fermions would annihilate at high pressures due to the restoration of TRS in a space-inversion protected system. A TPT could thus be expected along with the magnetic phase transition.

In order to study the possible pressure-induced magnetic and TPTs, we investigated the magnetic and electronic properties of $Co_3Sn_2S_2$ under pressures by using density function theory (DFT) calculations (See details in Methods). DFT results show that under ambient pressure, $Co_3Sn_2S_2$ is a FM nodal-line semimetal without spin-orbital coupling (SOC) [18-19, 42], where six line-nodes in the first Brillouin zone (BZ) near the Fermi level are formed by band inversion between two spin-up bands. The gaped band crosses are along the U-L and L-Γ paths (red bands in Figure 4(d) and also see Supplemental Materials Note 4.1). Similar crosses exist in conduction bands, which are at higher energy for the spin-down-dominant channel (blue bands in Figure 4(d)). The band structures under 0 GPa with and without SOC are almost the same except that the SOC lifts the band crossings along the U-L-Γ $k$-path.

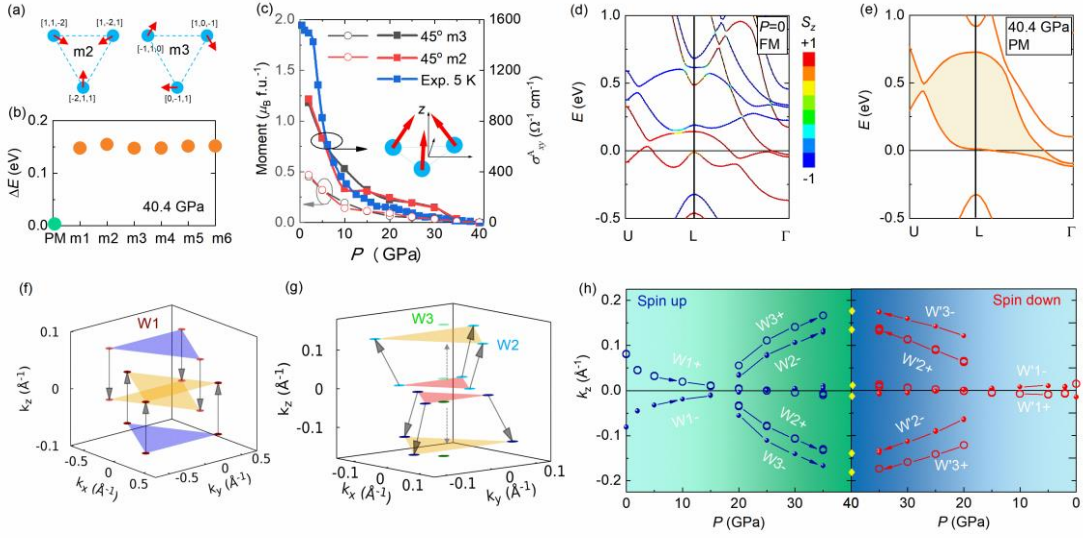

**Figure 4.** Theoretical calculations of magnetic and topological properties under pressure. (a) Two of considered in-plane AFM components. (b) Energy difference between the zero-moment PM state and six constrained in-plane AFM components (as shown in Supplemental Materials Note 4.2) with fixed moment directions and values (details in Supplemental Materials Note 4.4). The energy of the zero-moment PM state is set to zero. (c) Pressure dependences of the magnetic moment and AHC under two considered in-plane AFM m2 (red) and m3 (gray) configurations with fixed directions and relaxed values. The blue curve shows the measured pressure dependence of AHC. The inset sketches the tilted moments with out-of-plane FM and in-plane AFM components. The angle between moments and the $z$ axis should be varied during pressure loading. (d–e) Electronic structures of 0 (d) and 40 GPa (e) with high-symmetry paths in the Brillouin zone. The calculations were performed based on the configuration of out-of-plane linear ferromagnetic and PM structures, respectively. SOC is included and the color scale in (d) denotes the projection of the spin onto the $z$ axis. The Weyl nodes annihilate during the phase transition at $P_{c2}$ = 40.4 GPa. Thus, with an inverted band gap, a pressure-induced "topological insulator" phase emerges under both of inversion symmetry and TRS. (f), (g) Position evolutions of Weyl nodes W1 and W2 & W3 in spin-up channel. W1 exists during the whole loading, while W2 & W3 appear above 20 GPa. (h) The $k_z$-direction trajectory of the Weyl nodes from spin-up channel (blue color) and their counterparts from spin-down channel (red color), indicating an interband annihilation process near $P_{c2}$ = 40.4 GPa. Circles and squares denote Weyl nodes with chirality = +1 and -1, respectively.

Based on the above analyses about the possible in-plane AFM components, we performed two types of calculations. One type only considers out-of-plane linear FM and another includes in-plane AFM components in addition. The results indicate the later one is more consistent with the experimental results. For the in-plane AFM components, there are six considered magnetic structures with different in-plane AFM components, two of which are shown in Figure 4(a) (more details in Supplemental Materials Note 4.2). The calculated AHC with total moments tilting 45° away from the $z$ axis and the in-plane components order as m2 and m3 are shown in Figure 4(c) (right axis). Although the tilted angle should vary during the pressure increasing, our calculated results with fixed angle show same trend with the experimental data. Both experimental and theoretical (Figure 4(c)) results show that the AHC keeps fast decreasing with increasing pressure. Calculated total moments under the involved magnetic configurations are also shown in Figure 4(c). Moment starts to drop as soon as the pressure is loaded. Finally, the local moment of $Co_3Sn_2S_2$ vanishes just under the pressure around $P_{c2}$ = 40.4 GPa. Meanwhile, as Figure 4(e) shows, the bands of spin-up and spin-down channels degenerate at 40.4 GPa (see also Supplemental Materials Note 4.3), leading to zero local moment (no spin polarization) on Co atoms and thus, resulting in the restoration of TRS of the system. In an optical spectra research on $Co_3Sn_2S_2$, the temperature-dependent optical conductivity also reveals that the spin band splitting is reduced upon decreasing exchange interaction when the temperature goes upwards to $T_C$. It thus results in the reduction of both Berry curvature and AHC[43]. One can come to the conclude that the evolution of AHC directly correlates with the vanishing of ferromagnetism with increasing pressure or temperature in $Co_3Sn_2S_2$.

To confirm the Pauli PM is a stable magnetic state under high pressures, we performed further calculations based on six magnetic structures. All results show a consistent convergence to Pauli PM for all considered AFM components under the high pressure of 40.4 GPa. Moreover, all the AFM systems possess higher energy (Figure 4(b)) compared to PM if we constrain the AFM structures (see also Supplemental Materials Note 4.4). These results indicate a TRS restoration under 40.4 GPa. Hence, high pressure can also serve as an experimentally tunable 'knob' to switch the itinerant FM system to a TRS preserved Pauli PM phase. A TPT may occur concurrently here.

We now turn to the expected pressure-induced TPT. At relatively low pressures, we observe a Weyl phase transition with another two types of Weyl nodes (additional 4 pairs, denoted as W2 and W3) appear near pressure $P_{c1}$ = 20 GPa, which is similar to the previous report[28]. The position-evolution of these Weyl nodes is sketch in Figure 4(f) (for W1), 4(g) (for W2 and W3). Our DFT results further reveal a unique annihilation mechanism of the magneto-TPT. At high pressures, all the W1, W2 and W3 Weyl nodes (from spin-up channel) tend to move away from their partners with opposite chirality (see Figure 4(h) for $k_z$-trajectory and Supplemental Materials Note 4.5 for $k_x$- and $k_y$-trajectories), indicating that they do not annihilate intraband with each other at the magnetic transition point at 40.4 GPa. Instead, W1, W2 and W3 meet the corresponding band crossings W'1, W'2 and W'3 for the spin-down channel at 40.4 GPa. Therefore, the Weyl semimetal phase undergoes an interband annihilation and turns to a TPT. This is in sharp contrast to the nonmagnetic Weyl semimetal such as TaAs, in which the Weyl nodes with opposite chirality annihilate intraband with each other. Very recently, the TPT around the Curie temperature in $Co_3Sn_2S_2$ was revealed by temperature-dependent ARPES studies [15-17]. Previous and this studies indicate that this TPT in $Co_3Sn_2S_2$ can be driven by both temperature and pressure.

We furthurly show more details about the TPT. Figure 4(e) shows the band dispersions of PM $Co_3Sn_2S_2$ under pressures of 40.4 GPa. The annihilation of Weyl nodes is accompanied by the opening of a band gap (filled region). Therefore, as long as the band inversion persists, one can define a curved Fermi level within such a gap region and thus calculate the corresponding topological invariant. Our calculation by using Fu-Kane formula gives rise to a nontrivial $Z_2$ = 1, indicating that such high-pressure state is a "general TI phase" (where we relax the restriction of "a global band gap" to "a continuous band gap" between the highest valence band and the lowest conduction band). It means that the pressure up to 40.4 GPa can completely suppress the ground state of itinerant ferromagnetism and thus induce a TPT from magnetic Weyl semimetal to nonmagnetic TI phase with the annihilation of Weyl fermions in $Co_3Sn_2S_2$, as shown in Figure 1(c).

It is known that a strict TI contains two important ingredients, i.e., TRS and Fermi level lying inside the inverted bands gap. However, in an itinerant ferromagnetism system such as $Co_3Sn_2S_2$ we do not expect a global band gap. When the high pressure or temperature leads to a Pauli PM phase[36], the TRS is preserved both globally and locally. Together with the inversion symmetry, each energy band is at least double degenerate. However, the time-reversal symmetric $Co_3Sn_2S_2$ must have a half-filled band as the total electrons in its unit cell are an odd number. In contrast, when a magnetic Weyl semimetal dominated by local moments is tuned into a PM phase, the TRS is preserved globally but broken locally. One example is the recently discovered magnetic topological insulator $MnBi_2Te_4$[44], whose FM phase is predicted to be a magnetic Weyl semimetal[45-46] and PM phase is verified to be a three-dimensional TI due to the strong disorder of local moments[47]. Since the FM phase is not the ground state in $MnBi_2Te_4$, the TPT from magnetic Weyl semimetal to TI can be hardly observed without magnetic field. Thus, our present work proposes a route to realize magneto-TPT by tuning single knobs such as pressure or temperature. While pressure induced TPT can occur at very low temperatures, one can expect less thermal related influence and a more intrinsic transition process.

To summarize, a magneto-TPT from ferromagnetic Weyl phase to paramagnetic TI phase by external pressure was studied in a magnetic Weyl semimetal. Pressure drives the ground FM Weyl state to a new in-plane AFM and out-of-plane FM state with more Weyl nodes, and finally to the TRS preserved $Z_2$ TI phase. This coupling of magnetic-order and TPT offers an ideal arena for the emergent topological quantum phenomena. It will inspire more studies related to magneto-TPT in other material systems, e.g., with an intrinsic global gap in its nonmagnetic phase. We expect these evolutions of topological states in magnetic Weyl semimetals to extend the investigations of topological electronic states that manifest the exotic quantum states and suggest the potential for advancing topological electronics.

## Methods

**Growth and characterizations of single-crystalline microflakes.** All chemicals were of reagent-grade quality and used as received from commercial sources. The single crystal microflakes of $Co_3Sn_2S_2$ were grown by chemical vapor transport (CVT) method. We used the polycrystalline powders of $Co_3Sn_2S_2$ as source material and $I_2$ as a transport agent. In order to reduce the nucleation points during the growth process, a reverse temperature gradient was utilized before the crystal growth. The temperatures of the source and growth zones were first stabilized at 800 and 900°C, respectively, for 48 hours. Then the sample growth was initiated by setting the source and growth zones at 850 and 800°C, respectively, for two weeks. Hexagonal single-crystalline microflakes with dimension of 2−4 mm and thickness of around 10 μm were obtained. The magnetism, electric resistivity, and chemical composition of the microflakes were characterized, which confirmed the high-quality $Co_3Sn_2S_2$ samples. More details can be found in Supplemental Materials Note 1.

***In situ* high-pressure X-ray diffraction (XRD) measurements.** For the *in situ* high pressure XRD measurements, a symmetrical DAC with a pair of 400 μm diameter culet sized anvils was employed to generate high pressure. A Re gasket was pre-indented to 40 μm in thickness followed by laser-drilling to form a 150 μm diameter hole serving as the sample chamber. A well-grinded and pressed $Co_3Sn_2S_2$ pellet was loaded in the sample chamber. Silicone oil was used as the pressure-transmitting medium and the pressure was calibrated by using the ruby fluorescence shift at room temperature. The *in situ* synchrotron XRD experiments were carried out at the 4W1 High Pressure Station in the Beijing Synchrotron Radiation Facility (BSRF). The diffraction data were recorded by a Pilatus detector, and high purity $CeO_2$ powder was used for calibration.

**High-pressure electric transport measurements.** The high-pressure electric transport measurements were conducted on a $Co_3Sn_2S_2$ single crystal using the Quantum Design PPMS-9. A Be/Cu DAC with 300 μm culet anvils and a c-BN gasket were used. Four Pt wires were adhered to a $Co_3Sn_2S_2$ single crystal using the silver epoxy. Silicone oil was used as the pressure-transmitting medium and the pressure was calibrated by using the ruby fluorescence shift at room temperature. For the Hall measurements, magnetic fields were along the crystallographic *c* direction. The pressure was applied up to 40.3 GPa until the ferromagnetic order disappeared.

**Analysis of electric transport.** The longitudinal resistivity was calculated by Van der Pauw method[48] (see Supplemental Materials Note 3.1 for details). In practical measurements, there are weak transverse voltage drops on longitudinal data and longitudinal voltage drops on transverse data due to the mismatch of the electrodes. In order to obtain the accurate resistivity data, the procedure of (anti-)symmetrization was applied to extract the odd (even) parts of the observed signal. As the Hall curves show an observable nonlinear behavior in high magnetic field region between 10 to 30 GPa, which indicates the existence of two types of carriers (electrons and holes), two-band model analysis was thus employed[38, 49] after removing the AHE component. The details including carrier concentration and mobility can be found in Supplemental Materials Note 3.4.

**Density functional theory (DFT) calculations.** We applied DFT method by using the projector-augmented wave (PAW) pseudopotentials with the exchange-correlation of Perdew-Burke-Ernzerhof revised for solids (PBEsol) form and GGA approach as implemented in the Vienna ab-initio Simulation Package (VASP)[50-53]. For each pressure case, the crystal lattice and atomic position were fully relaxed until the atomic force on each atom is less than $10^{-2}$ eV/Å. We constructed Wannier representations by projecting the Bloch states from the DFT calculations of bulk materials onto Co-3$d$, S-3$p$, Sn-5$s$ and Sn-5$p$ orbitals. The AHC is expressed as the sum of the Berry curvature over all the occupied bands $\sigma_{xy}^{AH} = \frac{e^2}{\hbar} \frac{1}{N_k \Omega_c} \sum_k \sum_n f_{nk} \Omega_n^z(\boldsymbol{k})$, where $N_k, \Omega_c, f_{nk}$ and $\Omega_n^z(\boldsymbol{k})$ are total k-points number in BZ, volume of primitive cell, Femi distribution function and Berry curvature, respectively [54]. The locations and chiralities of Weyl nodes, AHC (with a 200×200×200 K-mesh), $Z_2$ number, and band/energy-occupied Berry curvatures were all calculated in the tight-binding models constructed by these Wannier representations[55-57], as implemented in the Wannier Tools package[8].